\documentclass[10pt,journal,compsoc]{IEEEtran}
\ifCLASSOPTIONcompsoc
  \usepackage[nocompress]{cite}
\else
  \usepackage{cite}
\fi
\ifCLASSINFOpdf
\else
\fi

\makeatletter
\def\ps@headings{%
\def\@oddhead{\mbox{}\scriptsize\rightmark \hfil \thepage}%
\def\@evenhead{\scriptsize\thepage \hfil \leftmark\mbox{}}%
\def\@oddfoot{}%
\def\@evenfoot{}}
\makeatother
\usepackage{subfigure}
\usepackage{graphicx, amssymb, amsbsy, amsmath, amsthm}
\usepackage{multirow}
\usepackage{threeparttable}
\usepackage{cite,url}

\usepackage{booktabs}
\usepackage{color,soul}
\usepackage[font=footnotesize,labelfont=sf,textfont=sf]{caption}
\usepackage[lined,ruled,linesnumbered]{algorithm2e}
\usepackage{algorithmic}
\usepackage{bbding}
\usepackage[colorlinks,linkcolor=red,anchorcolor=blue,citecolor=blue]{hyperref}
\graphicspath{{./figures/}}
\DeclareGraphicsExtensions{.pdf}
\UseRawInputEncoding

\def\squareforqed{\hbox{\rlap{$\sqcap$}$\sqcup$}}
\def\qed{\ifmmode\squareforqed\else{\unskip\nobreak\hfil
\penalty50\hskip1em\null\nobreak\hfil\squareforqed
\parfillskip=0pt\finalhyphendemerits=0\endgraf}\fi}

\begin{document}
%
\title{\emph{Opara:} Exploiting Operator Parallelism for Expediting DNN Inference on GPUs}
%
%
%
%

\author{Aodong~Chen,
        Fei~Xu,~\IEEEmembership{Member,~IEEE},
        Li~Han,~\IEEEmembership{Member,~IEEE},
        Yuan~Dong,
		Li~Chen,~\IEEEmembership{Member,~IEEE},
        Zhi~Zhou,~\IEEEmembership{Member,~IEEE},
        Fangming~Liu,~\IEEEmembership{Senior Member,~IEEE}
\IEEEcompsocitemizethanks{\IEEEcompsocthanksitem Aodong Chen, Fei Xu, and Yuan Dong are with the Shanghai Key Laboratory of Multidimensional Information Processing, School of Computer Science and Technology, East China Normal University, 3663 N. Zhongshan Road, Shanghai 200062, China. Email: {\tt fxu@cs.ecnu.edu.cn}.
\IEEEcompsocthanksitem Li Han is with the School of Software Engineering, East China Normal University, 3663 N. Zhongshan Road, Shanghai 200062, China. E-mail: {\tt hanli@sei.ecnu.edu.cn}.
\IEEEcompsocthanksitem Li Chen is with the School of Computing and Informatics, University of Louisiana at Lafayette, 301 East Lewis Street, Lafayette, LA 70504, USA. E-mail: {\tt li.chen@louisiana.edu}.
\IEEEcompsocthanksitem Zhi Zhou is with the Guangdong Key Laboratory of Big Data Analysis and Processing, School of Computer Science and Engineering, Sun Yat-sen University, 132 E. Waihuan Road, Guangzhou 510006, China. E-mail: {\tt zhouzhi9@mail.sysu.edu.cn}.
\IEEEcompsocthanksitem Fangming Liu is with Peng Cheng Laboratory, and Huazhong University of Science and Technology, China. E-mail: {\tt fangminghk@gmail.com}.
}
\thanks{Manuscript received August XX, 2023; revised May XX, 2024.}}

%
%

\markboth{Submitted to IEEE TRANSACTIONS ON COMPUTERS}%
{Chen \MakeLowercase{\textit{et al.}}: \emph{Opara:} Exploiting Operator Parallelism for Expediting DNN Inference on GPUs}
\IEEEtitleabstractindextext{%
\begin{abstract}

GPUs have become the \emph{defacto} hardware devices for accelerating Deep Neural Network (DNN) inference workloads. However, the conventional \emph{sequential execution mode of DNN operators} in mainstream deep learning frameworks cannot fully utilize GPU resources, even with the operator fusion enabled, due to the increasing complexity of model structures and a greater diversity of operators. Moreover, the \emph{inadequate operator launch order} in parallelized execution scenarios can lead to GPU resource wastage and unexpected performance interference among operators. In this paper, we propose \emph{Opara}, a resource- and interference-aware DNN \underline{Op}erator \underline{para}llel scheduling framework to accelerate DNN inference on GPUs. Specifically, \emph{Opara} first employs \texttt{CUDA Streams} and \texttt{CUDA Graph} to \emph{parallelize} the execution of multiple operators automatically. To further expedite DNN inference, \emph{Opara} leverages the resource demands of operators to judiciously adjust the operator launch order on GPUs, overlapping the execution of compute-intensive and memory-intensive operators. We implement and open source a prototype of \emph{Opara} based on PyTorch in a \emph{non-intrusive} manner. Extensive prototype experiments with representative DNN and Transformer-based models demonstrate that \emph{Opara} outperforms the default sequential \texttt{CUDA Graph} in PyTorch and the state-of-the-art operator parallelism systems by up to $1.68\times$ and $1.29\times$, respectively, yet with acceptable runtime overhead.

\end{abstract}

\begin{IEEEkeywords}
DNN inference, DNN operator parallelism, scheduling, GPU resource utilization
\end{IEEEkeywords}}

\maketitle

\IEEEdisplaynontitleabstractindextext

%
\IEEEpeerreviewmaketitle

\IEEEraisesectionheading{\section{Introduction}\label{sec:intro}}

\IEEEPARstart{D}{eep} Neural Networks (DNNs) have gained notable success in various business fields such as image processing, speech recognition, and virtual reality~\cite{pouyanfar2018survey}. In general, DNN inference tasks are exceptionally latency-sensitive. For instance, latency requirements in autonomous driving scenarios are non-negotiable (\emph{e.g.,} within $100$ milliseconds) due to safety considerations~\cite{mendoza2024model}. Accordingly, increasing attention from both academia and industry has been paid to efficient model serving. To meet such performance requirements, modern cloud datacenters are hosting thousands of GPUs to accelerate DNN inference for users. For instance, Alibaba Cloud houses more than $6,000$ GPUs, many of which are tasked with managing a substantial volume of inference requests~\cite{weng2022mlaas}.

Cloud-based GPUs are equipped with an increasing amount of computational power, which typically exceeds the resource demands of individual inference tasks, leading to under-utilization and wastage of hardware resources~\cite{li2022miso}. To achieve the objective model accuracy with fewer computations, several recent works (\emph{e.g.,} Szegedy et al.~\cite{szegedy2016rethinking}) focus on substituting large operators with several smaller and multiple-branch operators in DNN models, which further exacerbates the issue of GPU under-utilization. While batching requests~\cite{cui2022dvabatch} or co-locating model inference tasks~\cite{dhakal2020gslice} on a GPU can mitigate such GPU under-utilization, it inevitably prolongs the model inference due to batching latency and performance interference~\cite{xu2023igniter}. Moreover, the operator fusion~\cite{ma2020rammer} cannot fully utilize the GPU resources (as discussed in Sec.~\ref{sec:motivation-operator}) due to the limited scope of pre-defined fusion rules. Fortunately, as DNN models can typically be represented by a Directed Acyclic Graph (DAG) with \emph{parallel operators}, it provides us an opportunity to \emph{exploit inter-operator parallelism for accelerating DNN inference on GPUs while improving the GPU utilization.}

However, it is \emph{nontrivial} to efficiently parallelize the execution of DNN operators for a DNN inference task due to the following two facts. \emph{First,} the model DAG typically exhibits considerable complexity, often incorporating hundreds of operators with complex inter-operator dependencies. For simplicity, existing deep learning (DL) frameworks execute DNN operators one by one in topological sorting order~\cite{zhao2023autograph}. To achieve operator parallelism, a recent work (\emph{i.e.,} Nimble~\cite{kwon2020nimble}) relies on a reduction transformation of the DNN computation graph, which inevitably brings heavy computation overhead. \emph{Second,} inadequate operator parallel scheduling can adversely impact the DNN inference performance. As evidenced by motivation experiments in Sec.~\ref{sec:motivation-scheduling}, the inadequate operator launch order in mainstream DL frameworks (\emph{e.g.,} PyTorch) can prolong the DNN inference latency by up to $29\%$, due to the GPU blocking caused by the non-preemption feature of \texttt{CUDA} kernels~\cite{amert2017gpu} and performance interference among parallelizable operators~\cite{xu2023igniter}. In addition, several existing works (\emph{e.g.,} IOS~\cite{ding2021ios}) fail to consider the operator launch and function call overhead due to excessive CPU-GPU interactions when parallelizing DNN operators in the DL framework.



To address the challenges above, in this paper, we design \emph{Opara}, a resource- and interference-aware DNN \underline{Op}erator \underline{para}llel scheduling framework, with the aim of expediting the execution of DNN inference while improving the GPU utilization. We make the following contributions as below.

$\vartriangleright$ We propose a lightweight stream allocation algorithm without any modifications or transformations of the computation graph. It greedily allocates operators without dependencies to multiple \texttt{CUDA Streams} to maximize operator parallelism. Meanwhile, operators with data dependencies are allocated to the same \texttt{CUDA Stream} without impacting parallel executions of operators, thereby reducing the number of time-consuming synchronization operations.

$\vartriangleright$ We devise a resource- and interference-aware operator launch algorithm to judiciously prioritize launching operators with a small amount of GPU resource demands, so as to effectively mitigate GPU resource fragmentation and performance interference while reducing DNN inference latency. Such resource demands of operators can be obtained by lightweight inference profiling in practice.

$\vartriangleright$ We have implemented a prototype of \emph{Opara} (\url{https://github.com/icloud-ecnu/Opara}) as a plug-in module of PyTorch 2.0 to parallelize the executions of DNN operators. It can generate a parallelized \texttt{CUDA Graph} by capturing the stream allocation plan and optimized operator launch order to mitigate the operator launch and function call overhead. Our prototype experiments with $6$ representative DNN and Transformer-based models demonstrate that \emph{Opara} outperforms the default sequential \texttt{CUDA Graph} in PyTorch and the state-of-the-art DNN operator parallelism systems by up to $1.68\times$ and $1.29\times$, respectively.


\section{Background and Motivation}
\label{sec:motivation}

In this section, we first introduce how DNN operators are executed in mainstream DL frameworks, and identify the key factors that cause the low GPU utilization when serving DNN inference on GPUs. We then conduct motivation experiments to show how to judiciously parallelize the operator executions on GPUs.

\subsection{DNN Operator Executions on NVIDIA GPUs}
\label{sec:motivation-background}

After being scheduled on GPUs, a DNN operator is actually recognized as a kernel. In general, a kernel comprises multiple thread blocks, which are the smallest scheduling granularity in \texttt{CUDA}. A thread block is scheduled to a Streaming Multiprocessor (SM) once the SM has sufficient resources to meet its resource demands~\cite{gilman2022characterizing}. In particular, an SM can concurrently execute multiple thread blocks, and each SM is constrained by a limited number of threads, shared memory, and registers.

To enable parallel executions of operators, we launch operators on multiple \texttt{CUDA Streams}. Each stream is actually a task \emph{queue} that executes tasks sequentially. The execution order of kernels in different \texttt{CUDA Streams} is determined by their arrival order at the stream head. In general, the kernel execution time is considerably short as the batch size is typically small (\emph{i.e.,} ranging from $1$ to $16$) in latency-critical inference scenarios~\cite{strati2024orion, ma2020rammer}. Accordingly, the kernel launch overhead constitutes the primary time cost for DNN inference, which negatively impacts the performance gains achieved by the parallel executions of kernels in multiple \texttt{CUDA Streams}. To reduce such overhead, \texttt{CUDA Graph} is a key feature introduced from \texttt{CUDA 10} that allows scheduling multiple DNN operators on a GPU device at a time.

\subsection{Low GPU Utilization Due to Sequential Execution of DNN Operators}
\label{sec:motivation-operator}

Mainstream DL frameworks execute DNN operators \emph{sequentially} in topological sorting order, which cannot fully utilize GPU resources. To illustrate that, we conduct motivation experiments using the stock PyTorch 2.0 and ONNX Runtime 1.12\footnote{\url{https://onnxruntime.ai/}} with the operator fusion~\cite{ma2020rammer} enabled. We serve three typical DNN inference models including GoogLeNet\footnote{\url{https://pytorch.org/hub/pytorch_vision_googlenet/}}, Inception-v3~\cite{szegedy2016rethinking}, and BERT\footnote{\url{https://huggingface.co/google-bert}} on both an NVIDIA A100-PCIE-40GB GPU and an NVIDIA RTX 2080 SUPER GPU. In particular, we adopt the SM efficiency measured using NVIDIA Nsight Compute CLI\footnote{\url{https://docs.nvidia.com/nsight-compute/NsightComputeCli/index.html}} to evaluate the GPU utilization.

\begin{figure}
	\begin{minipage}[t]{0.5\linewidth}
		\centering
		\includegraphics[width=1.04\textwidth]{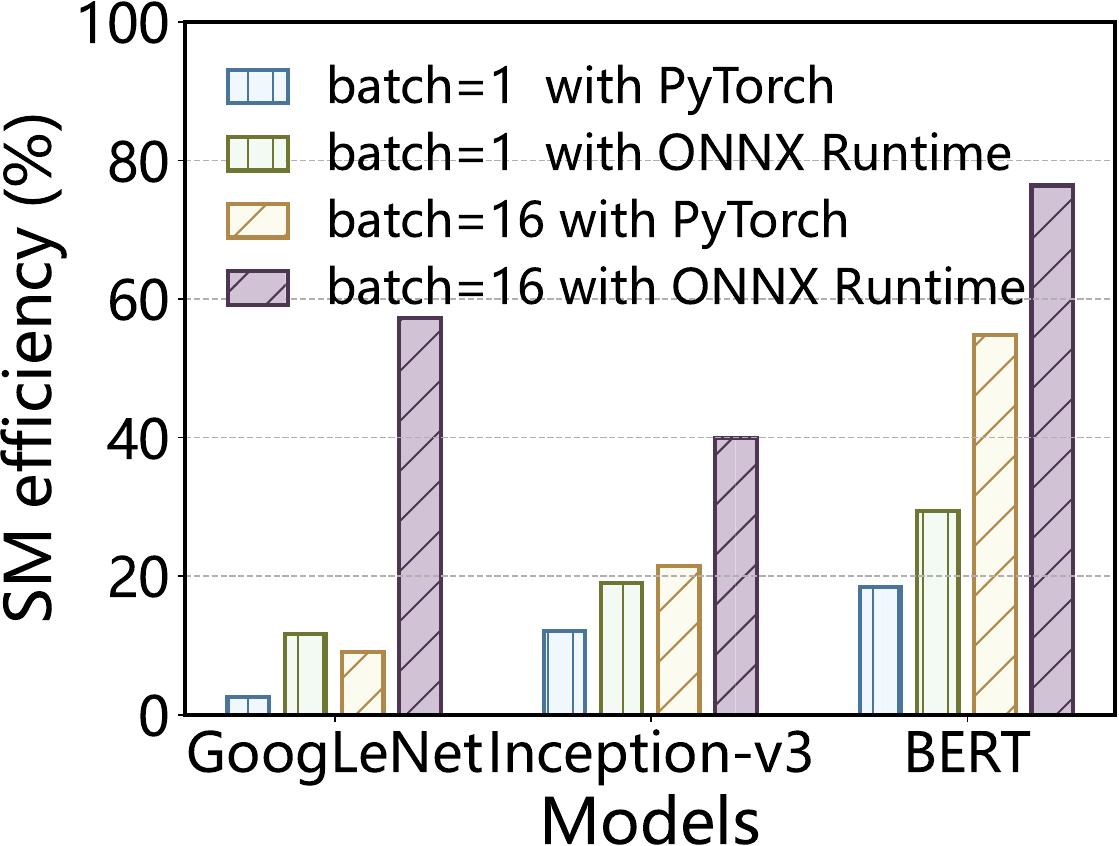}
        \caption{Average SM efficiency of an NVIDIA A100-PCIE-40GB GPU when running GoogLeNet, Inception-v3, and BERT models in PyTorch and ONNX Runtime.} \label{motivation_sm_efficiency_torch}
	\end{minipage}\hspace{+4pt}
	\begin{minipage}[t]{0.5\linewidth}
		\centering
		\includegraphics[width=0.96\textwidth]{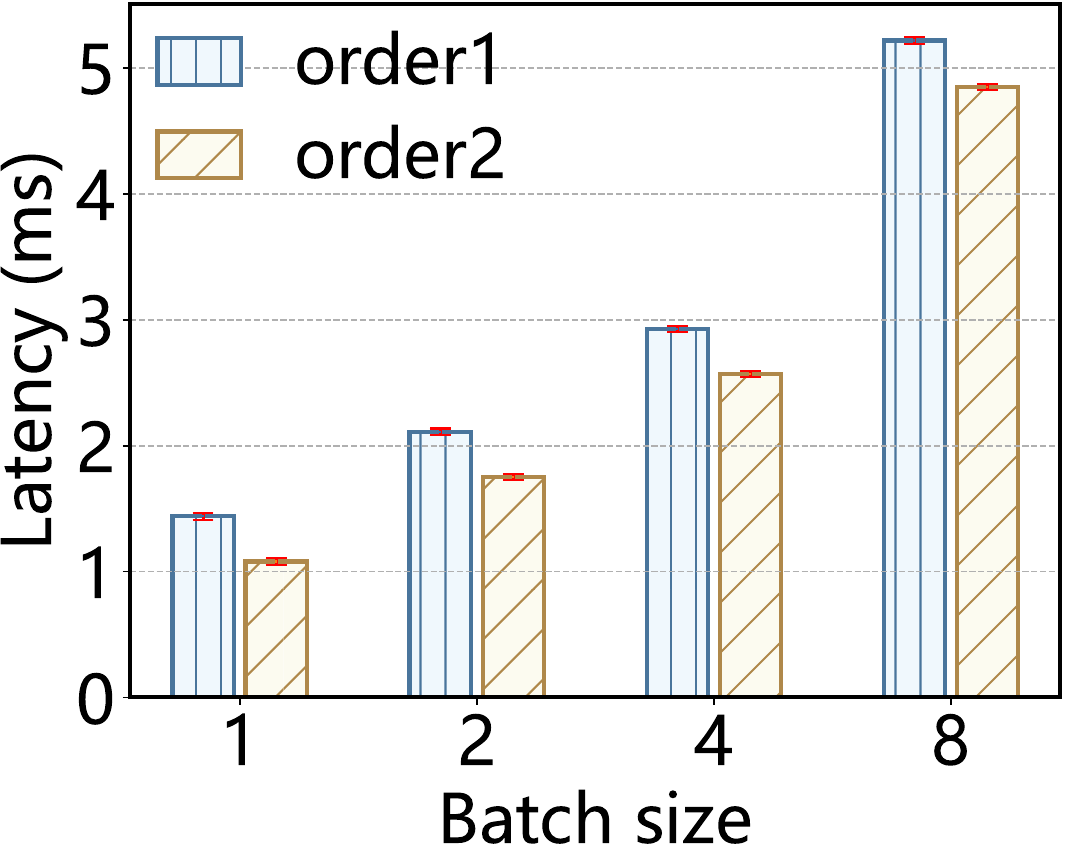}
        \caption{Inference latency of GoogLeNet running on an NVIDIA RTX 2080 Super GPU with different operator launch orders and batch sizes.} \label{motivation_launch_order}
	\end{minipage}\vspace{-10pt}
\end{figure}

As shown in Fig.~\ref{motivation_sm_efficiency_torch}, DNN inference on the mainstream DL frameworks achieves relatively low to medium GPU utilization even with the operator fusion technique enabled. Specifically, the SM efficiency of GoogLeNet, Inception-v3, and BERT with the batch size as $1$ is merely $2.53\%$, $12.04\%$, and $18.5\%$, respectively, achieved in the stock PyTorch on an A100 GPU. Even when the batch size is increased to $16$ and serving DNN inference in the ONNX Runtime, the SM efficiency of the three workloads is moderately increased to $57.21\%$, $39.9\%$, and $76.37\%$, respectively. Moreover, we repeat our experiments on a less powerful GPU (\emph{i.e.,} RTX 2080 SUPER), and the SM efficiency of the three workloads ranges from $10.47\%$ to $82.98\%$. Such experiment results above indicate that (1) the \emph{sequential} execution of DNN operators is the root cause of low GPU utilization for serving DNN inference; (2) The operator fusion technique can only combine a certain number of parallelizable operators based on the pre-defined fusion rules~\cite{ma2020rammer}, resulting in moderate GPU utilization. Accordingly, there still exists enough room to exploit the \emph{operator parallelism} for improving the GPU utilization of DNN inference on GPUs.


\subsection{Performance Impacts of Operator Launch Order}
\label{sec:motivation-scheduling}

Apart from the sequential execution of DNN operators, the \emph{inadequate operator launch order} (\emph{i.e.,} the topological sorting order of the model DAG) in mainstream DL frameworks can also lead to idle GPU resource usage and performance interference, thereby prolonging the inference latency. We conduct two motivation experiments to identify why the operator launch order can impact the inference latency.

\textbf{GPU blocking.} As each operator has a different number of blocks requiring three types of resources for execution, \emph{i.e.,} threads, shared memory, and registers, a \emph{resource-unaware} operator launch order can easily \emph{block} the execution of operators until enough resources become available on the GPU. Such \emph{GPU blocking} can severely waste the available GPU resources. As shown in Fig.~\ref{motivation_launch_order}, changing the operator launch order from order $1$ (\emph{i.e.,} depth-first topological sorting) to order $2$ (\emph{i.e.,} \emph{Opara} designed in Sec.~\ref{sec:design}) for GoogLeNet can reduce the inference latency by up to $29\%$ with different batch sizes. Furthermore, we repeat such an experiment on the A100 GPU, and the experiment results show around $10.3\%$ of performance improvement by optimizing the operator launch order for GoogLeNet.



\begin{figure}
\centering  
\centerline{\includegraphics[width=0.5\textwidth]{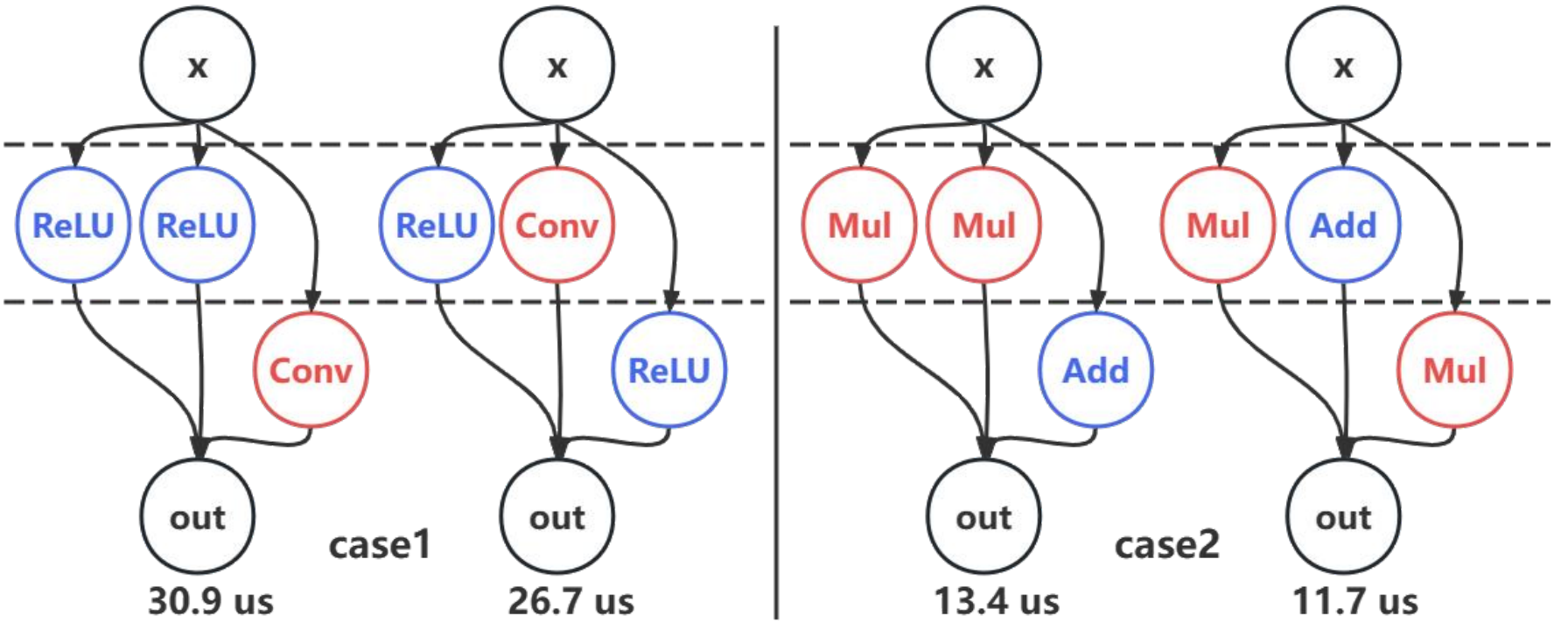}}
\caption{Overlapping the execution of compute-intensive and memory-intensive operators denoted as red circles and blue circles, respectively.} \label{motivation_compute_memory_normalized}
\vspace{-10pt}
\end{figure}

\textbf{Performance interference.} As the performance interference among operators can prolong the inference latency~\cite{xu2023igniter}, we further conduct another experiment on A100 to illustrate the effectiveness of \emph{overlapping the execution of compute-intensive and memory-intensive operators in mitigating the inference}.
As depicted in Fig.~\ref{motivation_compute_memory_normalized} (case 1), prioritizing the parallel execution of \texttt{ReLU} and \texttt{Conv} operators can cause less severe interference, compared with parallelizing two \texttt{ReLU} operators, leading to a $13.6\%$ reduction in the inference latency. Similarly, prioritizing the launch order of \texttt{Add} operator in case 2 can increase the inference performance by $12.7\%$, simply because the execution of compute-intensive and memory-intensive operators is overlapped.

\textbf{Summary.} Low GPU utilization of DNN inference is mainly caused by two factors: \emph{First}, the \emph{sequential} execution of DNN operators cannot fully utilize the GPU resources, even with the operator fusion enabled. \emph{Second}, the default topological sorting order of operator launch is commonly \emph{resource-} and \emph{interference-unaware}. Accordingly, judiciously parallelizing DNN operators with an adequate operator launch order is compelling for accelerating DNN inference on GPUs while improving the GPU utilization.


\section{System Design}
\label{sec:design}

\begin{figure}
\centering  
\centerline{\includegraphics[width=0.5\textwidth]{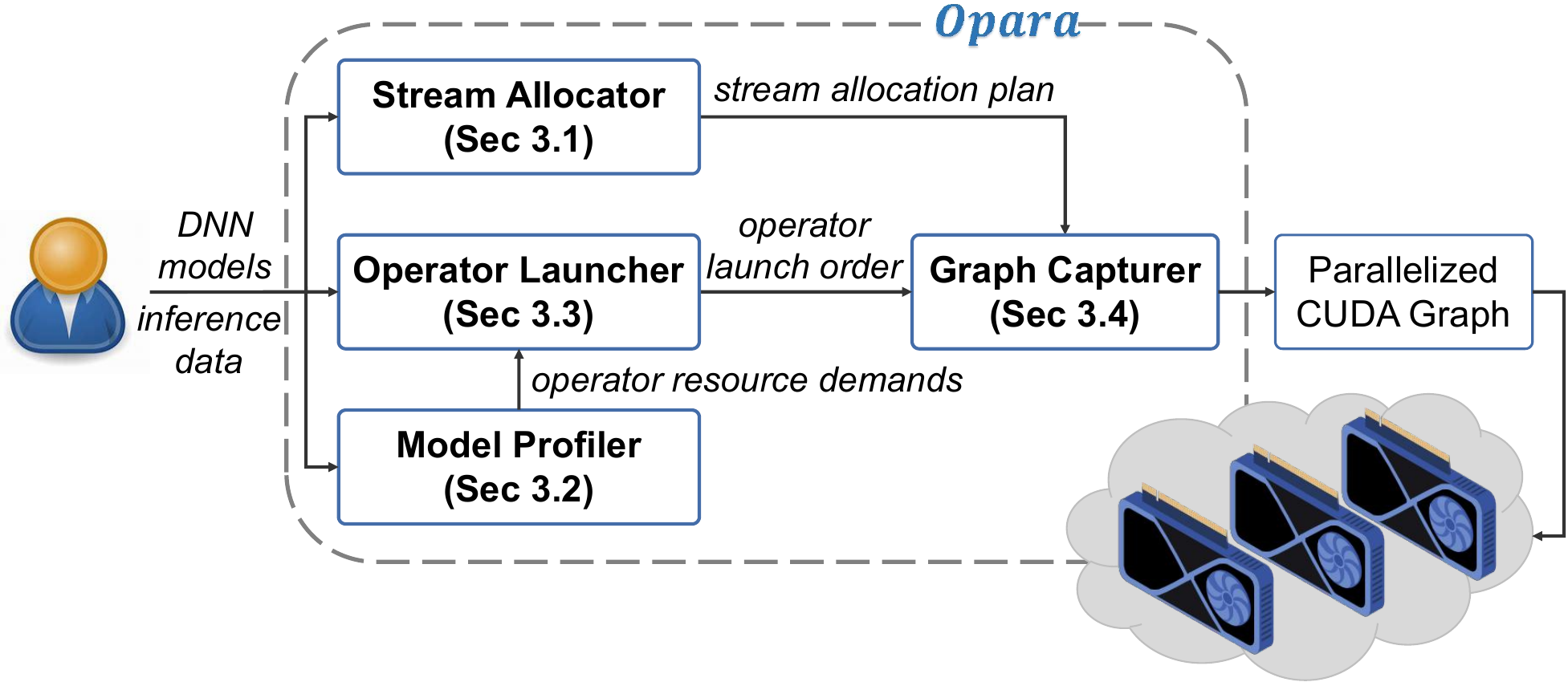}}
\caption{System overview of \emph{Opara}.} \label{overview}
\vspace{-10pt}
\end{figure}

In this section, we design \emph{Opara} illustrated in Fig.~\ref{overview}, an operator parallel scheduling framework to reduce DNN inference latency while improving the GPU resource utilization. Specifically, \emph{Opara} takes DNN models and input tensors (\emph{i.e.,} inference data) from users. According to the operator dependencies in the model DAG, the Stream Allocator first employs a stream allocation algorithm to determine which stream the operators should be allocated to. The Model Profiler then gathers the resource demands of each operator using the model profiling. With such resource demands of operators, the Operator Launcher further employs a resource- and interference-aware operator launch algorithm to optimize the operator launch order on GPUs. Finally, the Graph Capturer generates a parallelized \texttt{CUDA Graph} by combing the stream allocation plan and operator launch order, thereby enabling efficient DNN inference on GPUs.

\subsection{Stream Allocator}
\label{sec:design-allocator}

To parallelize the execution of operators in \texttt{CUDA Streams}, we leverage the computation graph (\emph{i.e.,} DAG) of DNN models to determine \emph{how many streams to launch and how to allocate operators to the streams.}

\textbf{Definition of a model DAG.} DNN computation graph can be represented as a DAG $\mathcal{G} = (\mathcal{V}, \mathcal{E})$, where $\mathcal{V}$ denotes the set of operators in the model, and $\mathcal{E}$ denotes the operator dependencies. Each vertex ${v} \in \mathcal{V}$ denotes a DNN operator (\emph{e.g.,} \texttt{Conv}, \texttt{MaxPool}). Each edge $\langle u, v\rangle \in \mathcal{E}$ denotes the operator dependency, where $u$ is a predecessor of ${v}$ and ${v}$ is a successor of ${u}$. The set of all predecessors of an operator ${v}$ are denoted as $\mathcal{N}_{pred}$. The set of all successors of an operator ${v}$ are denoted as $\mathcal{N}_{succ}$.

\textbf{Problem formulation and analysis.} As a maximum of $|\mathcal{V}|$ streams can be launched for a DAG, we simply use a matrix $\mathcal{A}$ of size $|\mathcal{V}| \times |\mathcal{V}|$ to represent the stream allocation plan, which determines how the DNN operators are parallelized and synchronized. Each element $a_{ij} \in \mathcal{A}$ is a boolean value, indicating whether the $i$-th operator is executed in the $j$-th stream. We formulate the inference latency $T_{inf}$ as
\begin{equation}\label{eq-inference}
    T_{inf} = T_{para} + T_{overhead},
\end{equation}
where $T_{para}$ denotes the parallelized execution time of a DNN model and $T_{overhead}$ denotes the operator synchronization overhead given a stream allocation plan $\mathcal{A}$. In more detail, as the plan $\mathcal{A}$ can parallelize the execution of DNN operators, $T_{para}$ can further be formulated as
\begin{equation}\label{eq-inference-parallel}
    T_{para} = h(\mathcal{A}) \times T_{seq},
\end{equation}
where $T_{seq}$ denotes the inference latency of \emph{sequential execution} of DNN operators and $h(\mathcal{A}) \in (0, 1]$ denotes the inference acceleration factor achieved by operator parallelism with the plan $\mathcal{A}$. To ensure the parallelized execution of the model, a certain number of synchronization operators~\cite{ng2023paella} require to be inserted into the model DAG. Such a process introduces significant operator synchronization overhead $T_{overhead}$, which can be formulated as
\begin{equation}\label{eq-inference-overhead}
    T_{overhead} = g(\mathcal{A}) \times t_{overhead},
\end{equation}
where $t_{overhead}$ is the time overhead caused by one synchronization operator and $g(\mathcal{A})$ is \emph{positively correlated} with the number of synchronization operators in approximation.

By substituting Eq.~(\ref{eq-inference-overhead}) and Eq.~(\ref{eq-inference-parallel}) into Eq.~(\ref{eq-inference}), we aim to minimize the inference latency $T_{inf}$ by identifying an optimal stream allocation plan $\mathcal{A}$. Accordingly, we formulate the stream allocation optimization problem as
\begin{eqnarray}
    \min_{\mathcal{A}} & & T_{inf} = h(\mathcal{A}) \times T_{seq} + g(\mathcal{A}) \times t_{overhead} \label{eq-optimal} \\
    \text { s.t. } & & \sum_{j=1}^{|\mathcal{V}|} a_{ij} = 1, \qquad \forall i \leq |\mathcal{V}| \label{cons-operator-placement}
\end{eqnarray}
where Constraint $(\ref{cons-operator-placement})$ mandates that each operator must be allocated to and only to one stream. $t_{overhead}$ and $T_{seq}$ can be considered as constant values given a DNN model. Actually, our optimization problem in Eq.~(\ref{eq-optimal}) (\emph{i.e.,} minimizing $h(\mathcal{A})$ and $g(\mathcal{A})$) can be considered as scheduling DAGs with dependency constraints (\emph{i.e.,} adjusting the matrix $\mathcal{A}$) to minimize the makespan, which has been proven to be an NP-hard problem~\cite{ullman1975np}. Accordingly, we turn to devising a heuristic algorithm to acquire an appropriate (\emph{i.e.,} sub-optimal) solution to our stream allocation problem.

\SetAlFnt{\small}
\SetAlgoVlined \vspace{-0pt}
\begin{algorithm}[!h]
\caption{Stream allocation algorithm in \emph{Opara}.}
\label{algorithm_1}
\SetAlgoNlRelativeSize{0}
\SetKwInOut{Input}{Input}
\SetKwInOut{Output}{Output}
\Input{DNN computation graph $\mathcal{G} = (\mathcal{V}, \mathcal{E})$}
\Output{Set of streams to be launched $\mathcal{S}$}
\BlankLine
\textbf{Initialize:} $\mathcal{S} \leftarrow \emptyset$, $\mathtt{SYNC}$ $flag$ $\gets \mathtt{False}$ for each operator $v \in \mathcal{V}$, and sort $\mathcal{V}$ in topological sorting order\;
\For{each operator $v \in \mathcal{V}$}{
    \For{each predecessor $p \in \mathcal{N}_{pred}$ of $v$}{
        \If{$flag$ of $p$ is $\mathtt{False}$}{
            $stream$ of $v \leftarrow$ $stream$ of $p$; \textcolor{blue}{\tcp{put $v$ and $p$ in the same stream}}
            $flag$ of $p \leftarrow \mathtt{True}$\;
            \textbf{break} out of the loop\;
        }
    }
    \If{$stream$ of $v$ is $null$}{
        $stream$ of $v \leftarrow$ launching a stream; \textcolor{blue}{\tcp{put $v$ in a newly launched stream}}
        $\mathcal{S} \leftarrow \mathcal{S} \cup \left\{stream \text{ of } v \right\}$\;
    }
}
\Return $\mathcal{S}$\;
\end{algorithm}

\textbf{Stream allocation algorithm.} The \emph{key idea} of Alg.~\ref{algorithm_1} is to allocate parallelizable operators to multiple \texttt{CUDA Streams} as much as possible (\emph{i.e.,} minimizing the value of $h(\mathcal{A})$). Moreover, we greedily put non-root nodes (\emph{i.e.,} operators) in the same \texttt{CUDA stream} as one of their predecessor operators, so as to avoid excessive synchronization operators (\emph{i.e.,} minimizing the value of $g(\mathcal{A})$). Specifically, given a computation graph $\mathcal{G}$, \emph{Opara} first initializes a set of streams to be launched $\mathcal{S}$ and the $\mathtt{SYNC}$ $flag$ of each $v \in \mathcal{V}$. It then enumerates operators in $\mathcal{V}$ in topological sorting order (lines $1$-$2$). For each operator ${v} \in \mathcal{V}$, it iterates over all of its predecessors $p \in \mathcal{N}_{pred}$ (line $3$). If the current predecessor $p$ has not yet contributed to reducing the synchronization overhead (\emph{i.e.,} $flag$ is $\mathtt{False}$), it allocates $v$ to the same stream of $p$, and set the $flag$ of $p$ as $\mathtt{True}$ (lines $4$-$7$). If $v$ does not find a predecessor that satisfies such a condition above, we allocate the operator $v$ to a newly launched stream (lines $8$-$10$). In particular, the parallelized execution of streams does not impact each other as long as operators are not executed on GPUs. To ease the understanding of Alg.~\ref{algorithm_1}, we present an illustrative example in Appendix~\ref{sec:appendix-example}.

\subsection{Model Profiler}
\label{sec:design-profiler}

As discussed in Sec.~\ref{sec:motivation-scheduling}, the blocks in an operator execute the same instructions even with different data, which indicates that the GPU resources required by the blocks in an operator are the same. Accordingly, we obtain the resource demands of each operator by simply profiling the resource consumption (\emph{i.e.,} the amount of shared memory, the number of registers and GPU threads) of a block in an operator. Such resource demands of operators will be used by Operator Launcher to determine an adequate operator launch order. In particular, we implement our Model Profiler utilizing the \texttt{torch.profiler.profile()} API, and it requires profiling each DNN inference \emph{only once} to acquire the resource demands information for each operator, thereby bringing acceptable profiling overhead. We will examine the inference profiling overhead of \emph{Opara} in Sec.~\ref{sec:evaluation-overhead}.

\subsection{Operator Launcher}
\label{sec:design-launcher}

\textbf{Problem analysis.} As illustrated in Sec.~\ref{sec:motivation-scheduling}, inadequate operator launch orders can significantly affect the DNN inference latency. To identify an optimal launch order, a naive solution is iterating through \emph{all possible} topological sorting orders of a model DAG and choosing the order with the lowest inference latency. However, such a method involves selecting nodes with zero indegree and deleting the corresponding vertices and their connected edges. By assuming $n$ operators exist in a model DAG, the time complexity of traversing all topological sorting orders is $\mathcal{O}(n!)$, which is also an NP-hard problem~\cite{ullman1975np}. As a result, we turn to designing a heuristic operator launch algorithm to solve such a complex problem.

\begin{algorithm}[!h]
\caption{Operator launch algorithm in \emph{Opara}.}
\label{algorithm_2}
\SetAlgoNlRelativeSize{0}
\SetKwInOut{Input}{Input}
\SetKwInOut{Output}{Output}
\Input{DNN computation graph $\mathcal{G} = (\mathcal{V}, \mathcal{E})$}
\Output{Operator queue $\mathcal{Q}$ in resource- and interference-aware operator launch order}
\BlankLine
\textbf{Initialize:} List of operators to be launched $\mathcal{L} \leftarrow \emptyset$, $\mathcal{L}_{mem} \leftarrow \emptyset$, $\mathcal{L}_{comp} \leftarrow \emptyset$, and $\mathcal{Q} \leftarrow \emptyset$\;
Add the operators $v \in \mathcal{V}$ with an $indegree$ of $0$ that are \emph{memory-intensive} and \emph{compute-intensive} to $\mathcal{L}_{mem}$ and $\mathcal{L}_{comp}$, respectively\;
\While{$\mathcal{L}_{mem}$ or $\mathcal{L}_{comp}$ is not empty}{
    $\mathcal{L} \gets$ alternately choose a non-empty list from ${\{\mathcal{L}_{mem}, \mathcal{L}_{comp}\}}$\;
    $v_{min} \gets$ the operator that requires the least amount of GPU resources in $\mathcal{L}$\; 
    $\mathcal{L}.{remove}({v_{min}})$, and $\mathcal{Q}.{append}({v_{min}})$; \textcolor{blue}{\tcp{launch the operator $v_{min}$}}
    \For{each successor $s \in \mathcal{N}_{succ}$ of $v_{min}$}{
        Update $indegree$ of ${s} \gets$ $indegree$ of ${s} - 1$\;
        \If{$indegree$ of ${s} == 0$}{
            \If{operator ${s}$ is \emph{memory-intensive}}{
                $\mathcal{L}_{mem}.\text{append}(s)$; \textcolor{blue}{\tcp{add $s$ to $\mathcal{L}_{mem}$}}
            }\Else{
                $\mathcal{L}_{comp}.\text{append}(s)$; \textcolor{blue}{\tcp{add $s$ to $\mathcal{L}_{comp}$}}
            }
        }
    }
}
\Return{$\mathcal{Q}$}\;
\end{algorithm}

\textbf{Resource- and interference-aware operator launch algorithm.} Launching operators with heavy resource demands first to the GPU is likely to cause resource fragmentation, hindering the GPU executions of subsequent operators. Moreover, the GPU can thus be blocked due to the non-preemptive feature of kernel execution~\cite{amert2017gpu}. To maximize the GPU utilization, the \emph{key idea} of Alg.~\ref{algorithm_2} is to greedily prioritize launching the operators with the least amount of GPU resource demands, aiming at maximizing the parallel executions of multiple operators within a model. To further mitigate the performance interference among operators~\cite{xu2023igniter}, we simply \emph{overlap} the execution of compute-intensive operators and memory-intensive operators, as classified by our offline-collected operator table.

Specifically, it first initializes and maintains a priority queue $\mathcal{Q}$ of operators in resource- and interference-aware operator launch order (line $1$). It then retrieves all operators to be launched with an indegree of $0$ in a list $\mathcal{L}$, which alternates between the non-empty lists for memory-intensive operators $\mathcal{L}_{mem}$ and for compute-intensive operators $\mathcal{L}_{comp}$ (lines $2$-$4$). Each time the operator requiring the least amount of GPU resources (\emph{e.g.,} shared memory, threads, registers) is chosen from $\mathcal{L}$ and then put into the queue $\mathcal{Q}$ (lines $5$-$6$). In particular, the potential GPU blocking issue faced by the remaining large operators is \emph{noncritical} in our scenario, as $\mathcal{L}$ is dynamic and can be compensated for the upcoming small operators to be launched. Finally, $\mathcal{L}_{mem}$ and $\mathcal{L}_{comp}$ are continuously updated by adding new operators with an indegree of $0$ (lines $7$-$13$).

\subsection{Graph Capturer}
\label{sec:design-capturer}


To eliminate the overhead caused by kernel launches and function calls, the Graph Capturer first sets the \texttt{CUDA Streams} obtained from the Stream Allocator to the capture mode, and then it launches the operators of the DNN model to these streams according to the operator launch order specified by the Operator Launcher. To ensure the dependencies among operators, the Graph Capturer also launches the necessary synchronization operators to the streams. Consequently, a \texttt{CUDA Graph} is generated to enable operator parallelization while improving the GPU utilization. Such a graph capture process is lightweight and non-intrusive to PyTorch, as it has been exposed as a high-level API in PyTorch officially. We simply use the PyTorch API to capture and then generate the \texttt{CUDA Graph}.


\section{Implementation of \emph{Opara}}
\label{sec:implementation}

We implement a prototype of \emph{Opara} with around $1,000$ lines of Python codes, which have been integrated into PyTorch 2.0 as a plug-in module. The source codes are currently publicly available on GitHub (\url{https://github.com/icloud-ecnu/Opara}). Specifically, we employ \texttt{torch.fx.Graph} as the computation graph for DNN models in \emph{Opara}. Its Intermediate Representation (IR) allows us to schedule DNN operators directly in Python. In more detail, we leverage the \texttt{torch.cuda.set\_stream()} API in PyTorch to launch operators on the \texttt{CUDA Streams}. To particularly guarantee the operator dependency in parallelized executions of streams, we add the appropriate synchronization operators to the model graph using the \texttt{event.record()} and \texttt{stream.wait\_event(event)} APIs. Finally, we use \texttt{torch.cuda.graph(g)} to generate a \texttt{CUDA Graph} that can execute DNN operators in parallel based on the \texttt{CUDA Streams}. In summary, we build our prototype of \emph{Opara} only using the high-level APIs of PyTorch in a lightweight and non-intrusive manner, rather than modifying the computation graph construction module as in Nimble~\cite{kwon2020nimble}.

\section{Performance Evaluation}
\label{sec:evaluation}

In this section, we carry out prototype experiments to demonstrate the efficacy and runtime overhead of \emph{Opara} in comparison to the stock PyTorch and state-of-the-art operator parallelism frameworks.

\subsection{Experimental Setup}
\label{sec:evaluation-setup}

\textbf{Hardware configuration and workloads.} We conduct our experiments on an NVIDIA A100-PCIe-40GB GPU and an NVIDIA GeForce RTX 2080 SUPER-8GB GPU. We implement \emph{Opara} based on CUDA 11.7, cuDNN 8.5.0, and as a plug-in module of PyTorch 2.0. Our experiments employ $6$ representative DNN models, including Inception-v3~\cite{szegedy2016rethinking}, GoogLeNet, DeepFM~\cite{guo2017deepfm}, NASNet\footnote{\url{https://huggingface.co/timm/nasnetalarge.tf_in1k}}, BERT, and T5\footnote{\url{https://huggingface.co/google-t5}}. The first three models are executed on the RTX 2080 GPU, while the latter three are run on the A100 GPU.

\textbf{Baselines and metrics.} We compare DNN inference performance of \emph{Opara} with that of the stock PyTorch (with \texttt{CUDA Graph} disabled), default sequential \texttt{CUDA Graph}, ONNX Runtime (with operator fusion enabled), Rammer~\cite{ma2020rammer}, and Nimble~\cite{kwon2020nimble}. Specifically, we implement Rammer's BFS-based operator scheduling algorithm (named Wavefront) into PyTorch 2.0. Nimble transforms the computation graph into a bipartite graph and identifies its maximum matching to determine an appropriate stream for each operator. In particular, we focus on $4$ key metrics including DNN inference latency, SM efficiency, and GPU memory consumption, as well as DNN inference throughput. All the experiment results are averaged over $1,000$ runs.

\begin{figure}
    \centering
    \subfigure[]{\includegraphics[width=0.86\linewidth]{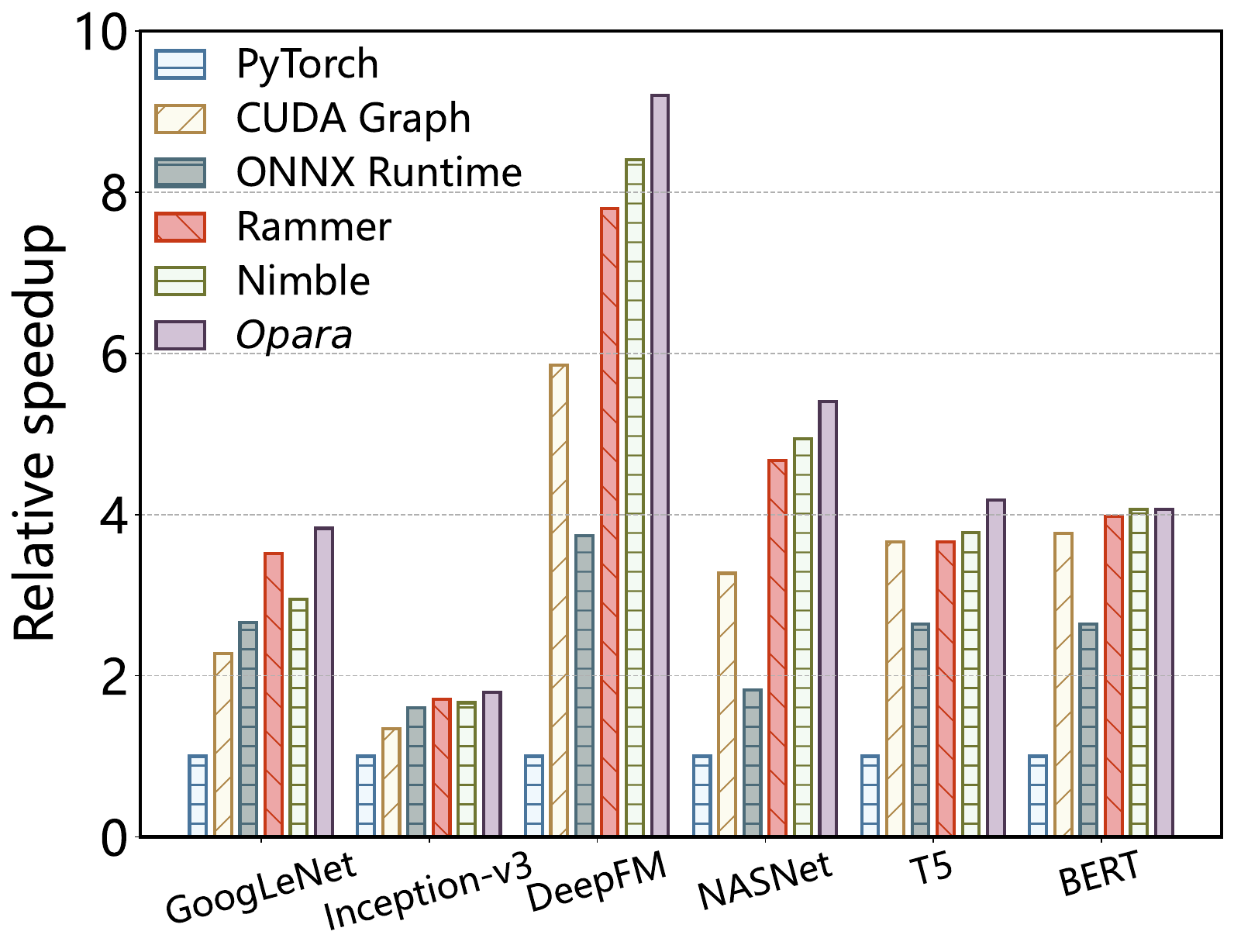}
    \label{relative_speedup_2080s_A100}}\hspace{0pt}\vspace{-8pt}
    \subfigure[]{\includegraphics[width=1\linewidth]{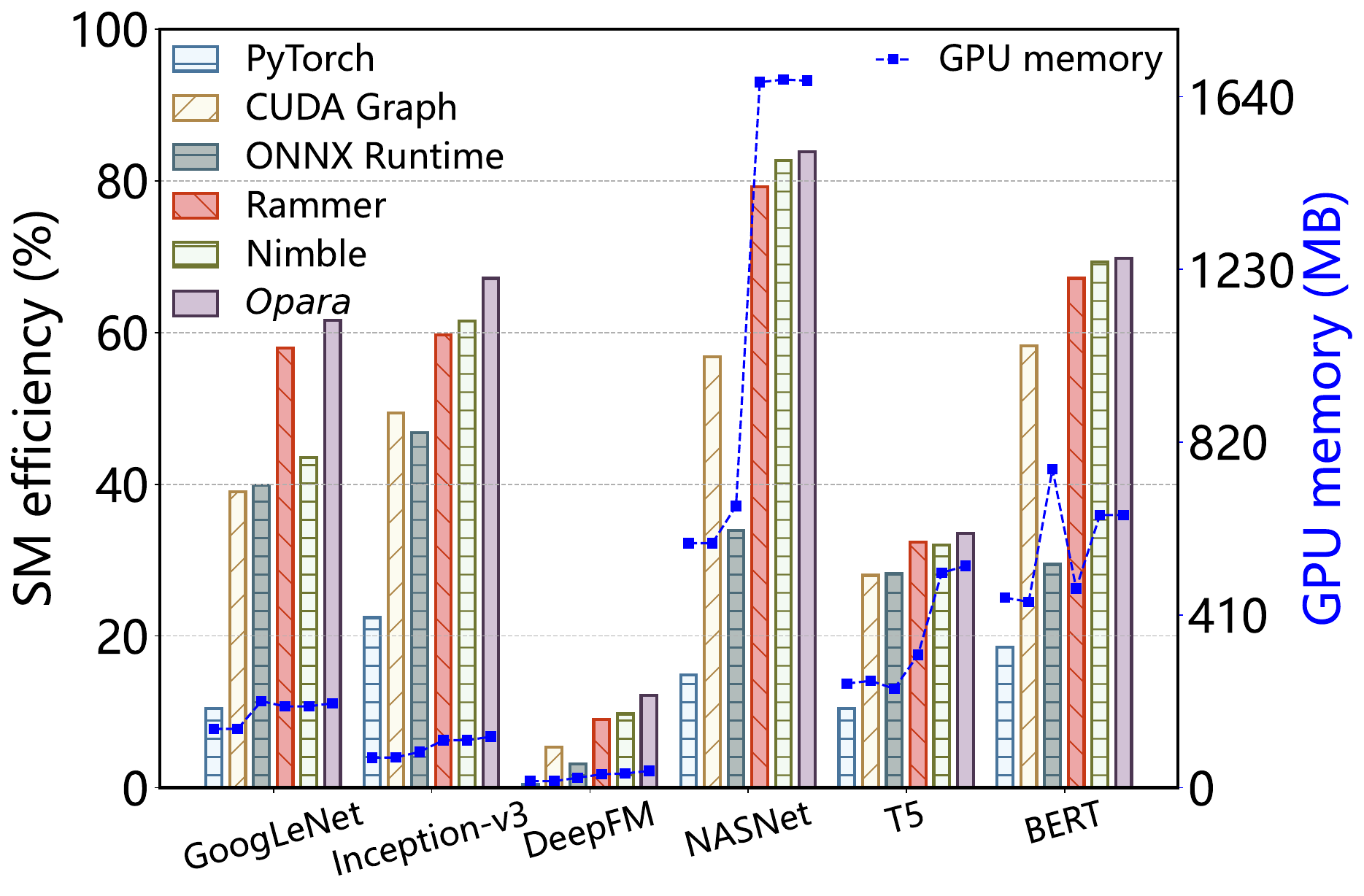}
    \label{sm_efficiency_2080s}}
    \vspace{-8pt}
    \caption{(a) Relative speedup, (b) SM efficiency and peak memory consumption of GPUs running representative DNN models with batch size set as $1$ achieved by PyTorch, \texttt{CUDA Graph}, ONNX Runtime, Rammer, Nimble, and \emph{Opara} operator scheduling mechanisms.}
    \vspace{-10pt}
\end{figure}

\subsection{Effectiveness of \emph{Opara}}
\label{sec:evaluation-effectiveness}

\textbf{End-to-end inference latency.} As shown in Fig.~\ref{relative_speedup_2080s_A100}, \emph{Opara} consistently outperforms the five baselines for six representative DNN models. Specifically, \emph{Opara} can achieve $1.80\times$ to $10.97\times$ speedup compared to the stock PyTorch. This is because \emph{Opara} utilizes \texttt{CUDA Graph} to eliminate the operator launch and function call overhead. \emph{Opara} surpasses the default \texttt{CUDA Graph} by up to $1.68 \times$, simply because of the parallel execution of DNN operators in \emph{Opara}. Though the operator fusion technique outperforms the stock PyTorch, \emph{Opara} achieves a higher speedup by up to $2.97\times$ and $1.18\times$, compared with ONNX Runtime and Rammer, respectively. Such performance improvements above are mainly due to two facts: \emph{First,} the operator fusion cannot combine all parallelizable operators based on the pre-defined fusion rules, which cannot fully utilize the GPU resources. \emph{Second,} \emph{Opara} accelerates model inference through operator parallelization, while the Wavefront scheduling algorithm in Rammer introduces additional synchronization overhead (\emph{i.e.,} the unnecessary operator waiting time during wave executions).
Furthermore, \emph{Opara} outperforms Nimble by up to $1.29 \times$ because it judiciously alternates the scheduling of different types of operators with the lowest GPU resource consumption for each kernel launch time. Moreover, \emph{Opara} initiates enough streams to increase parallelism (\emph{e.g.,} $28$ streams with \emph{Opara} versus $4$ streams with Nimble for GoogLeNet), thereby maximizing the operator parallelism. 

\textbf{GPU utilization and memory consumption.} To unveil the performance gains of \emph{Opara}, we proceed to look into the GPU utilization (\emph{i.e.,} SM efficiency) and memory consumption during the model inference. As shown in Fig.~\ref{sm_efficiency_2080s}, \emph{Opara} exhibits a similar improvement in GPU utilization compared to the five baselines as in Fig.~\ref{relative_speedup_2080s_A100}. Specifically, \emph{Opara} significantly improves the GPU utilization compared to the stock PyTorch, because \emph{Opara} mitigates the scheduling overhead of the stock PyTorch. When compared with the default \texttt{CUDA Graph}, \emph{Opara} increases the GPU utilization of Inception-v3, GoogLeNet, DeepFM, NASNet, BERT, and T5 by $36\%$, $58\%$, $126\%$, $48\%$, $20\%$, and $19\%$, respectively. Such performance gains mainly come from the parallelized execution of operators. When compared to ONNX Runtime, Rammer, and Nimble, \emph{Opara} boosts the GPU utilization by up to $3.86\times$, $1.36\times$ and $1.42 \times$ mainly because (1) maximizing stream allocations in \emph{Opara} can increase operator parallelism opportunities, and (2) optimizing the operator launch order in \emph{Opara} further minimizes the GPU idle time. Furthermore, the parallel execution of operators requires an increased amount of data to reside in the GPU memory simultaneously, thereby leading to a higher peak GPU memory consumption of \emph{Opara} than that of sequential executions.

\begin{figure}
    \centering
    \includegraphics[width=0.5\textwidth]{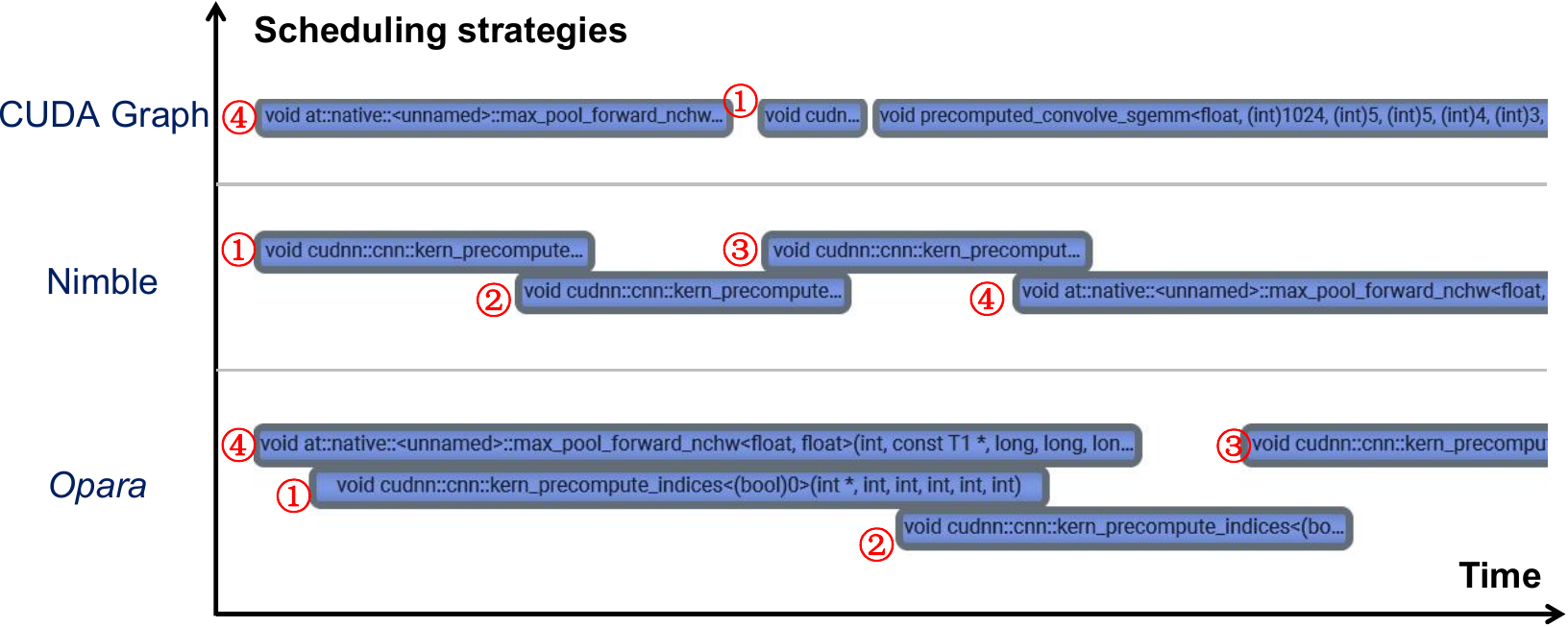}
    \caption{Timeline of operator executions during a segment of inference process of GoogLeNet achieved by \texttt{CUDA Graph}, Nimble, and \emph{Opara}.} \label{timeline}
    \vspace{-10pt}
\end{figure}

\textbf{Timeline of operator executions.} We further illustrate the operator execution timeline by taking a segment of inference process of GoogLeNet as an example. In particular, we leverage NVIDIA Nsight System CLI\footnote{\url{https://docs.nvidia.com/nsight-systems/UserGuide/index.html}} to track the timeline of operator executions. As depicted in Fig.~\ref{timeline}, we observe that \texttt{CUDA Graph} executes the $4$ operators sequentially in a stream, and only operators $4$ and $1$ appear within the time window. The remaining operators $2$ and $3$ are forced to queue up, which leads to a long inference time. Though Nimble can parallelize operators in the order of $1$, $2$, $3$, and $4$, it only schedules $2$ operators on two streams, causing a long GPU idle time. In contrast, \emph{Opara} prioritizes operator $4$ and initiates more streams than Nimble, so that operators $4$, $1$, and $2$ can be executed in parallel to maximize the operator parallelism. Accordingly, \emph{Opara} can achieve the shortest inference latency by exploring operator parallelism compared with \texttt{CUDA Graph} and Nimble.

\begin{figure}
    \centering
    \subfigure[T5]{\includegraphics[width=0.9\linewidth]{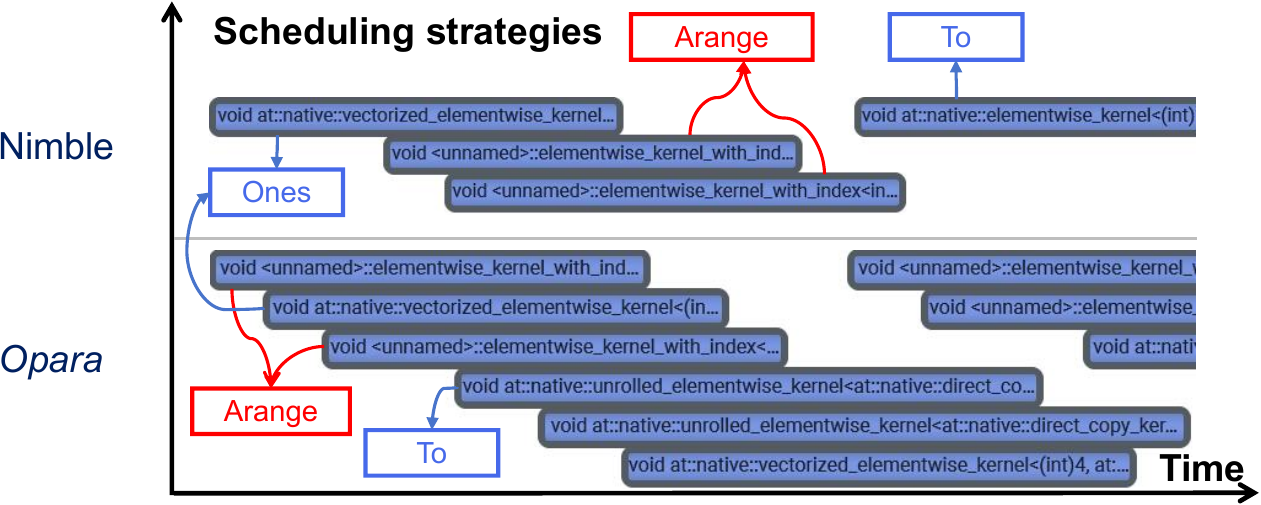}
    \label{timeline-of-T5}}\hspace{0pt}
    \subfigure[BERT]{\includegraphics[width=0.9\linewidth]{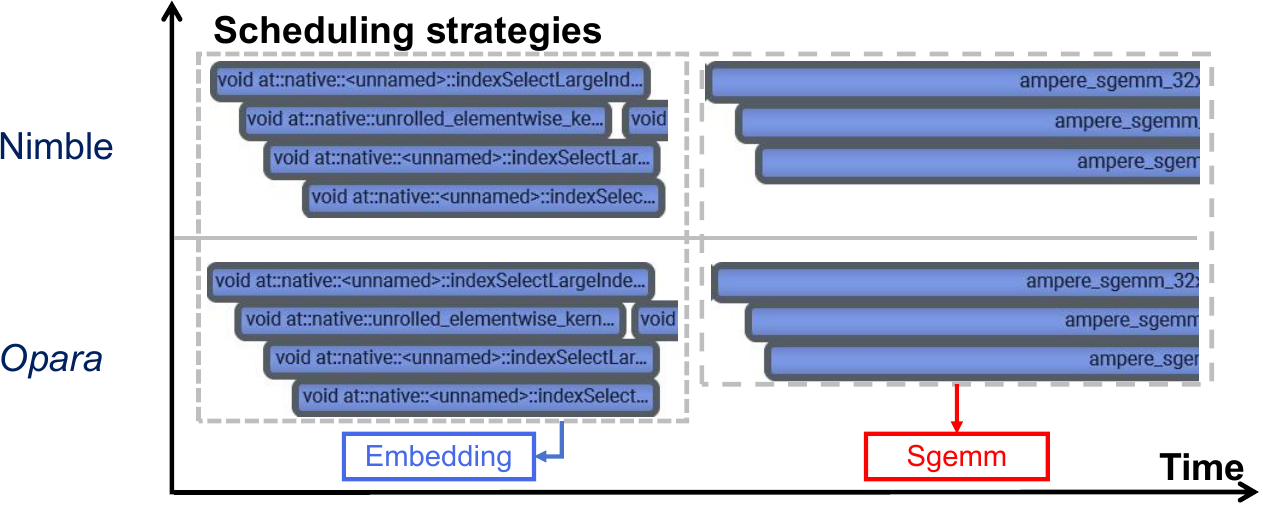}
    \label{timeline-of-BERT}}
    \vspace{-8pt}
    \caption{Timeline of Transformer-based models achieved by Nimble and \emph{Opara} operator scheduling mechanisms. Red rectangles and blue rectangles represent compute-intensive operators and memory-intensive operators, respectively.}
    \vspace{-10pt}
\end{figure}

\textbf{Effectiveness of \emph{Opara} on Transformer-based models.} We conduct experiments with T5 and BERT model, and \emph{Opara} outperforms Nimble by $9.3\%$ for the T5 model as shown in Fig.~\ref{relative_speedup_2080s_A100}. This is because \emph{Opara} optimizes the launch order of operators in T5 and schedules them into $6$ streams compared with $3$ streams in Nimble. Moreover, the \emph{operator diversity} in T5 offers \emph{Opara} overlap the compute-intensive \texttt{Arange} operators and the memory-intensive \texttt{To} and \texttt{Ones} operators, as shown in Fig.~\ref{timeline-of-T5}. For BERT, however, \emph{Opara} achieves a similar operator launch order and the same number of streams as Nimble as depicted in Fig.~\ref{timeline-of-BERT}. This is because the parallelizable operators of BERT are always the \texttt{Embedding} operators or the \texttt{Sgemm} operators, which reduces the opportunity for operator overlapping and launch order optimization. Accordingly, \emph{Opara} achieves marginal performance gains for BERT compared with Nimble, yet $1.08\times$ to $4.06\times$ speedup compared to the stock PyTorch and \texttt{CUDA Graph} as shown in Fig.~\ref{relative_speedup_2080s_A100}.

\begin{figure}
	\begin{minipage}[t]{0.5\linewidth}
		\centering
        \includegraphics[width=1.02\textwidth]{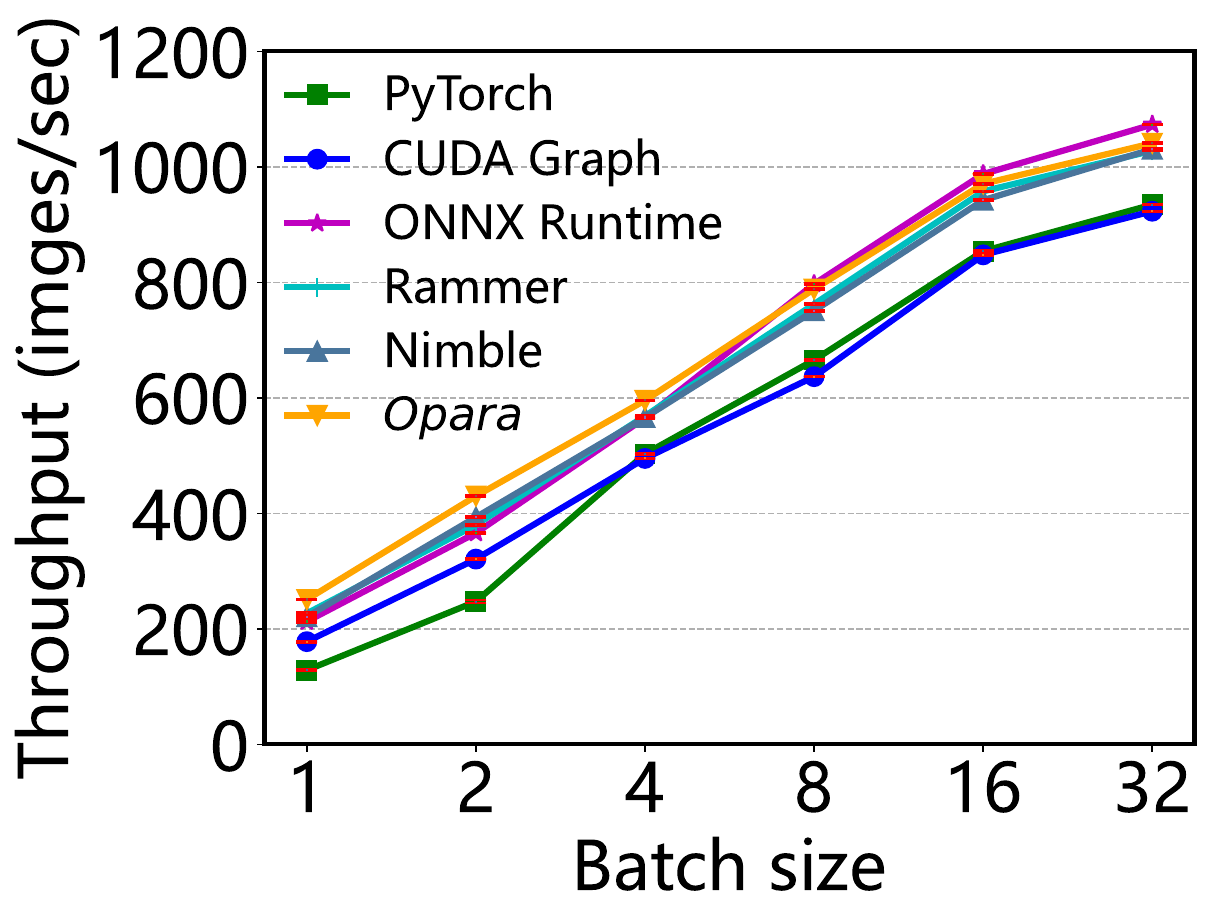}
        \caption{Inference throughput of Inception-v3 with \emph{Opara} and the five baselines by varying the batch size from $1$ to $32$ on an RTX 2080 GPU.} \label{diff_batch_throughput_2080s}
	\end{minipage}\hspace{+4pt}
	\begin{minipage}[t]{0.5\linewidth}
		\centering
		\includegraphics[width=0.92\textwidth]{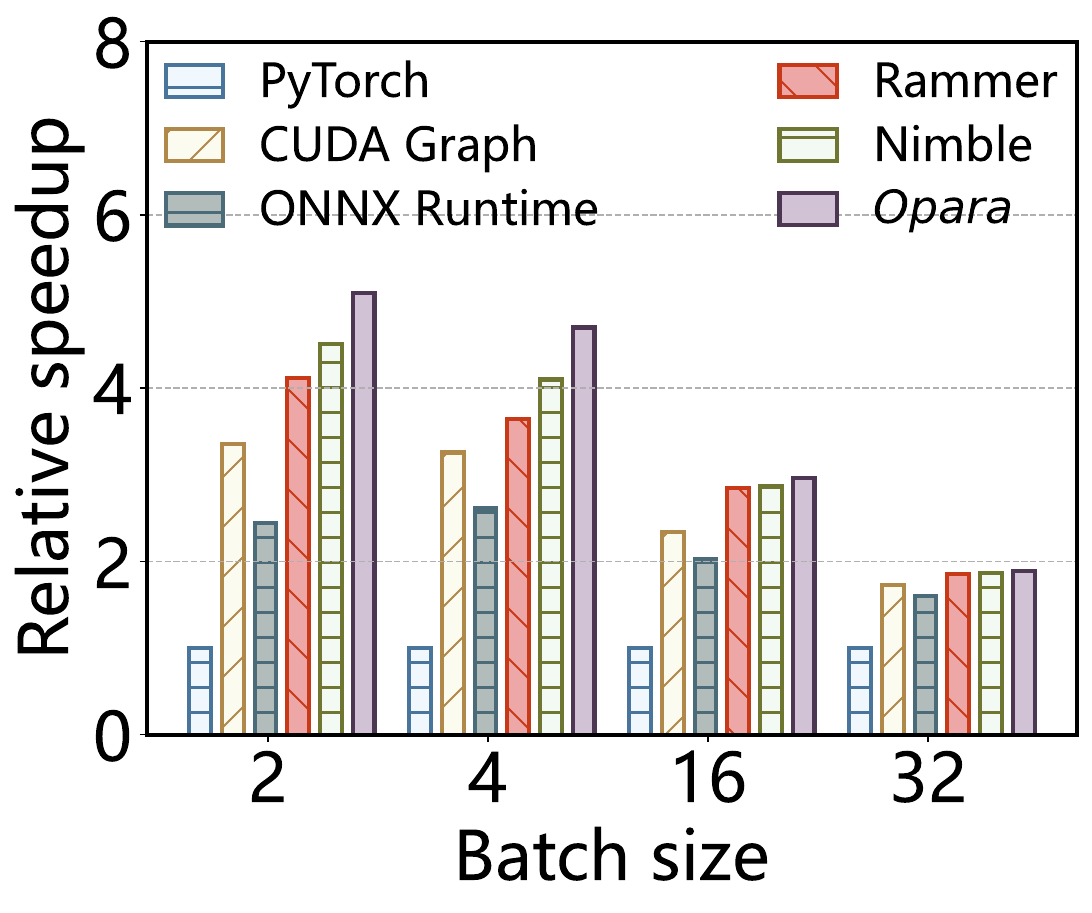}
        \caption{Relative speedup of Inception-v3 with \emph{Opara} and the five baselines by varying the batch size from $2$ to $32$ on an A100 GPU.} \label{relative_speedup_A100}
	\end{minipage}\vspace{-10pt}
\end{figure}

\textbf{Throughput under different batch sizes.} As depicted in Fig.~\ref{diff_batch_throughput_2080s}, we observe that \emph{Opara} consistently surpasses the five baselines except for ONNX Runtime by varying the batch size from $1$ to $32$. Nevertheless, the performance gains of \emph{Opara} gradually diminish as the batch size increases. As an example, \emph{Opara} outperforms the default \texttt{CUDA Graph} by $1.41 \times$ and $1.09 \times$ when the batch size is $1$ and $32$, respectively. This is because the amount of GPU resources occupied by a single operator increases when dealing with larger batch sizes, resulting in fewer GPU resources available for the execution of parallelized operators. This also explains why \emph{Opara} exhibits marginal throughput improvement for large batch sizes of $16$ and $32$. The results above also show that maximizing the operator parallelism can also improve the inference throughput.

\textbf{Effectiveness of \emph{Opara} on high-end GPUs with sufficient resources.} We repeat the inference experiment of Inception-v3 on a high-end GPU (\emph{i.e.,} A100). As shown in Fig.~\ref{relative_speedup_A100}, we observe that \emph{Opara} consistently outperforms the five baselines by varying the batch size from $2$ to $32$, mainly because operator parallelism works well for high-end GPUs with sufficient resources. In more detail, \emph{Opara} achieves an inference speedup by up to $2.08\times$, $1.29\times$, and $1.15\times$ compared to ONNX Runtime, Rammer, and Nimble, respectively. In particular, \emph{Opara} achieves speedups of $1.47\times$ and $1.18\times$ relative to ONNX Runtime for batch sizes of $16$ and $32$, which is larger than the results achieved on the RTX 2080 GPU. This is because the A100 GPU provides sufficient resources, which allows the operator parallelism to achieve more performance gains than the operator fusion.

\subsection{Runtime Overhead of \emph{Opara}}
\label{sec:evaluation-overhead}

We evaluate the runtime overhead of \emph{Opara} in terms of algorithm computation time and inference profiling overhead. As listed in Table~\ref{runtime_overhead}, \emph{Opara} can reduce the computation time of the stream allocation algorithm by up to two orders of magnitude compared with Nimble~\cite{kwon2020nimble}. This is because Nimble requires a graph transformation together with an exhaustive search in the bipartite graph. Such a process is time-consuming with a complexity in the order of $\mathcal{O}({n}^{3})$, where $n$ is the number of operators in a model DAG. In contrast, the time complexity of \emph{Opara} can be reduced to the order of $\mathcal{O}({n})$, simply because the inner loop of Alg.~\ref{algorithm_1} (lines $3$-$10$) in \emph{Opara} only depends on the maximum width (\emph{i.e.,} typically below $20$) of the computation graph. Accordingly, as DNN models become increasingly complex~\cite{shi2023welder}, the number of operators $n$ gets even larger, while the algorithm computation overhead of \emph{Opara} can still be well contained. In addition, as the Model Profiler needs to run the DNN inference only once, \emph{Opara} requires several (\emph{i.e.,} $4.25$) milliseconds of profiling overhead in our experiment. In sum, the runtime overhead of \emph{Opara} is practically acceptable.

\begin{table}
    \centering
    \caption{Computation time (in milliseconds) of the stream allocation algorithm in \emph{Opara} (\emph{i.e.,} Alg.~\ref{algorithm_1}) and Nimble~\cite{kwon2020nimble} for various models.} \label{runtime_overhead}
    \begin{tabular}{lccccc}
        \toprule
        & BERT & GoogLeNet & NASNet & Inception-v3 & T5\\
        \midrule
        \emph{Opara} & $0.58$ & $0.27$ & $1.75$ & $0.50$ & $2.8$ \\
        
        Nimble & $20.8$ & $5.80$  & $257.83$ & $14.40$ & $161.4$ \\
        \bottomrule
    \end{tabular}
    \vspace{-10pt}
\end{table}



\section{Related Work}
\label{sec:related}

\textbf{Inter-operator parallelism within a single model.} To parallelize the execution of DNN operators, Rammer~\cite{ma2020rammer} proposes fine-grained operator scheduling based on the Wavefront algorithm and enables operator fusion on a GPU device. 
To increase the operator parallelism, Cocktailer~\cite{zhang2023cocktailer} further co-schedules control flow and data flow operators based on Rammer. 
The two prior works above operate at the \emph{compilation} level, which requires significant compilation overhead and manual customization of operators. Orthogonal to them, \emph{Opara} focuses on the \emph{runtime operator scheduling} optimization of the stream allocation and the operator launch order. A recent work Nimble~\cite{kwon2020nimble} leverages the bipartite graph algorithm to schedule operators on \texttt{CUDA streams} adequately. IOS~\cite{ding2021ios} deploys operator fusion and dynamic programming to determine operator parallelization plans. However, Nimble and IOS require a lengthy search process and neglect the optimization space of operator launch order. In contrast, \emph{Opara} utilizes the \texttt{CUDA Graph} to eliminate such performance overhead. It also employs a \emph{lightweight} stream allocation algorithm to achieve inter-operator parallelism. To reduce the GPU idle time and interference, \emph{Opara} determines a feasible operator launch order according to operator resource demands.

\textbf{Inter-operator parallelism among different models.} To improve GPU utilization, several works parallelize operators from multiple models co-located on a GPU device. For example, S$^{3}$DNN~\cite{zhou2018s} and Abacus~\cite{cui2021enable} optimize the co-location of operators from different models and schedule them to the corresponding steams. 
To minimize the model co-location interference, \emph{iGniter}~\cite{xu2023igniter} and Orion~\cite{strati2024orion} focus on optimizing the GPU resource allocation and operator scheduling on multiple prioritized streams, respectively.
Paella~\cite{ng2023paella} dispatches the optimal kernel from multiple models by jointly considering the remaining time and model fairness. Different from optimizing the inference co-location, \emph{Opara} minimizes the inference latency while increasing the GPU utilization by parallelizing operators within a single model. Moreover, it achieves inter-operator parallelism as a plug-in module of PyTorch 2.0 without developing a new DL runtime or framework.

\textbf{Intra-operator parallelism.} Existing DL frameworks, such as PyTorch and TensorFlow, employ expert-optimized operator libraries to accelerate individual operators. TVM~\cite{chen2018tvm} uses machine learning methods to automatically search for efficient operators, which is time-consuming and requires the specified parameter space manually. To achieve automated code generation, Ansor~\cite{zheng2020ansor} implements an automatic search space construction. As a single DNN operator cannot fully utilize GPU resources in general, \emph{Opara} can work with the intra-operator parallelism methods above to further improve GPU resource utilization.

\vspace{-0pt}
\section{Conclusion and Future Work}
\label{sec:conclusion}

This paper presents the design and implementation of \emph{Opara}, a lightweight operator scheduling framework to speed up DNN inference on GPUs. By reducing the synchronization overhead among operators, \emph{Opara} designs a stream allocation algorithm to automatically allocate operators without dependencies to different \texttt{CUDA streams}, thereby achieving operator parallelism effectively. Furthermore, \emph{Opara} leverages non-intrusive inference profiling to judiciously select an appropriate operator launch order to mitigate interference and maximize the GPU utilization. Extensive prototype experiments show that \emph{Opara} can improve the performance of DNN inference by up to $29\%$, as compared to the state-of-the-art operator parallelism systems.

We plan to extend \emph{Opara} in the following directions: (1) constructing an analytical model to analyze the performance interference caused by inter-operator parallelism, and (2) examining the effectiveness of \emph{Opara} for accelerating more large models (\emph{e.g.,} GPT-3, LLaMA).

\ifCLASSOPTIONcaptionsoff
  \newpage
\fi

\vspace{-10pt}
\bibliographystyle{IEEEtran}
\bibliography{ref}

\newpage
\clearpage
\appendix

\subsection{An Illustrative Example of Alg.~\ref{algorithm_1}}
\label{sec:appendix-example}

\begin{figure}[hbt]
\centering
\includegraphics[width=0.34\textwidth]{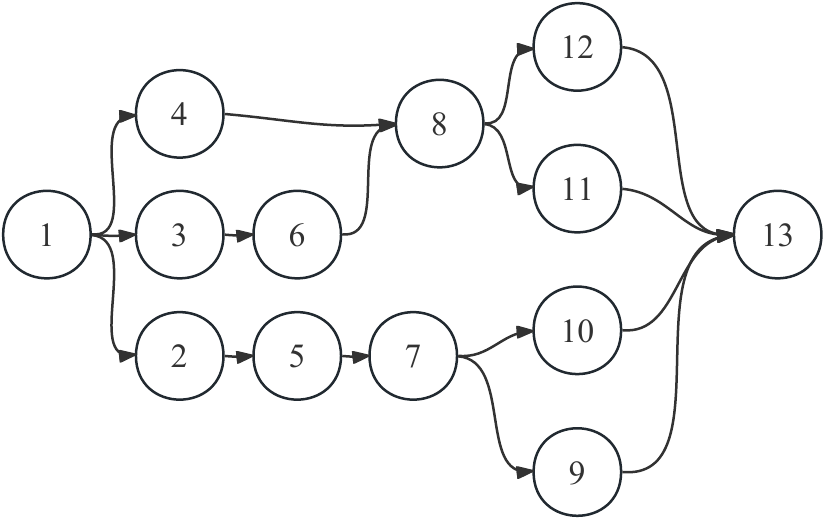}
\caption{A stream allocation example with a typical DNN model.} \label{dag}
\vspace{-10pt}
\end{figure}

As an illustrative example in Fig.~\ref{dag}, Alg.~\ref{algorithm_1} first traverses the operators in the computation graph in topological sorting order. It then categorizes the operators into four types based on their predecessor relationships as follows.

\begin{itemize}
    \item For an operator $i$ without a predecessor, like operator $1$ in the example, we place the operator $i$ in a newly-created stream.
    \item For an operator $i$ with only one predecessor and its predecessor has only one successor, like operators $5$, $6$, and $7$ in the example, we place the operator $i$ in the same stream as its predecessor.
    \item For an operator $i$ with a unique predecessor that has multiple successors, if the operator $i$ is the first successor of its predecessor, like operators $2$, $9$, and $11$ in the example, we place the operator $i$ in the same stream as the predecessor; otherwise, like operators $3$, $4$, $10$, and $12$ in the example, we place the operator $i$ in a newly created stream. This is because its predecessor has already contributed to reducing the synchronization overhead for the first successor (\emph{i.e.,} the $\mathtt{SYNC}$ $flag$ of their predecessors has been set as $\mathtt{True}$).
    \item For an operator $i$ with multiple predecessors, like operators $8$ and $13$ in the example, we place the operator $i$ in the same stream as the first predecessor with the $\mathtt{SYNC}$ $flag$ set as $\mathtt{False}$, by traversing all its predecessors.\\
\end{itemize}

\end{document}